\documentclass[11pt]{paper}

\expandafter\let\csname equation*\endcsname\relax
\expandafter\let\csname endequation*\endcsname\relax

\usepackage{lineno}
\usepackage{amsmath}
\usepackage{graphicx}% Include figure files
\usepackage{xfrac}
\usepackage{upgreek}
\usepackage{color}
\begin{document}

\title{Electromagnetically-induced transparency assists the Raman gradient echo memory at moderate detuning, dependent on gradient order}

\author{Jesse L. Everett*, Ankit Papneja, Arindam Saha,\\ Cameron Trainor, Aaron D. Tranter, Ben C. Buchler\\
\small{*jesse.everett@anu.edu.au}}

\maketitle
\section*{Abstract}
Optical quantum memories are essential for quantum communications and photonic quantum technologies. Ensemble optical memories based on 3-level interactions are a popular basis for implementing these memories. All such memories, however, suffer from loss due to scattering. In off-resonant 3-level interactions, such as the Raman gradient echo memory (GEM), scattering loss can be reduced by a large detuning from the intermediate state. In this work, we show how electromagnetically induced transparency adjacent to the Raman absorption line plays a crucial role in reducing scattering loss, so that maximum efficiency is in fact achieved at a moderate detuning. Furthermore, the effectiveness of the transparency, and therefore the efficiency of GEM, depends on the order in which gradients are applied to store and recall the light. We provide a theoretical analysis and show experimentally how the efficiency depends on gradient order and detuning.

\section{Introduction}
The development of optical quantum memories is motivated by use in quantum communication and networking \cite{briegel_quantum_1998,bussieres_prospective_2013,xia_long_2017, zapatero_advances_2023}, quantum sensing \cite{pirandola_advances_2018}, and for use directly in classical or quantum optical computing \cite{bussieres_prospective_2013, campbell_configurable_2014}.
Ensemble optical memories rely on a controllable, coherent interaction between light and a long-lived electronic state, with the light absorbed into and regenerated from a collective excitation of the ensemble of emitters \cite{choi_mapping_2008,duan_long-distance_2001, zhang_creation_2009}. Polarisation, time, frequency, and other degrees of freedom of the signal light are reversibly mapped into the spatial and internal degrees of freedom of the emitters \cite{hammerer_quantum_2010,heshami_quantum_2016}. A high efficiency of this mapping is necessary for preserving quantum states \cite{fleischhauer_quantum_2002,lvovsky_optical_2009}, and helpful for improving communication bandwidths in quantum repeaters \cite{jobez_towards_2016, simon_quantum_2007}. 

Examples of ensemble optical memory schemes include Electromagnetically-Induced Transparency (EIT) \cite{fleischhauer_electromagnetically_2005}, Atomic Frequency Comb (AFC) \cite{afzelius_demonstration_2010}, Raman \cite{le_gouet_raman_2009,wasilewski_pairwise_2006,reim_towards_2010}, Autler-Townes Splitting (ATS) \cite{saglamyurek_coherent_2018}, and Gradient Echo Memory (GEM) \cite{hetet_photon_2008}.
Several platforms are suitable for various selections of these schemes, including warm alkali atom vapors \cite{lukin_controlling_2001, hetet_photon_2008, katz_light_2018}, cold thermal atoms \cite{cho_highly_2016, hsiao_highly_2018}, and dopant atoms or colour centres in solids \cite{atature_material_2018,awschalom_quantum_2018,lago-rivera_telecom-heralded_2021,duda_optimizing_2023}. Ensemble memories hold records for the highest efficiency quantum memories \cite{cho_highly_2016,hsiao_highly_2018}.

In 3-level schemes such as EIT and GEM, the information in some `signal' light is mapped into the atomic ensemble with the help of some `control' light, as illustrated in the inset of Fig.~\ref{fig:concept}a). In EIT, the signal light is resonant with the transition to the excited state, and the interaction of the control field with the 3-level system opens a transparency window, preventing the incoherent scattering of the light. In GEM and other Raman type memories, the control field instead generates a Raman absorption peak near the 2-photon resonance, as illustrated in Fig.~\ref{fig:concept} a). Both these interactions generate a \textit{spinwave}, which is a coherent, collective excitation of the long-lived ground states  \cite{fleischhauer_quantum_2002}. To recover light from the memory, the control field must be reapplied to convert the spinwave coherence into a travelling optical field.
The optical control in 3-level schemes allows a significant degree of flexibility to control the timing, frequency, and bandwidth of the retrieved light \cite{finkelstein_fast_2018,vernaz-gris_high-performance_2018}.
These 3-level ensemble memories all have common features associated with the control field that govern their theoretical maximum efficiencies based on a limiting resource of optical depth \cite{gorshkov_photon_2007}.

Efficiency and lifetime are further limited by other properties of the ensemble, such as the motion of atoms or other sources of broadening, which scramble the phase of the emitters. Physics beyond the 3-level model can also limit the memory, for example giving lower fidelity due to added noise from four-wave mixing  \cite{hsiao_highly_2018}. Operating a memory that can preserve more degrees of freedom, or using the memory to process the stored information, will further limit the efficiency \cite{gorshkov_photon_2007}. This is due to the optical depth being divided over more inputs and outputs of the memory.

GEM \cite{hetet_photon_2008,moiseev_efficiency_2008}, illustrated in Fig.~\ref{fig:concept}, combines the off-resonant Raman interaction with a frequency gradient along the length of the atomic ensemble, typically a Zeeman shift due to a magnetic field gradient. This sets a bandwidth for the memory, where different frequency components of the signal pulse are absorbed into the atomic coherence at the locations of their respective 2-photon resonances. { A spatial winding up of the spinwave phase due to the frequency gradient prevents re-emission of the light from the coherence. Flipping the gradient unwinds the spinwave and the light is re-emitted, as long as the control laser is present.}

An efficiency of 87$\%$ was previously demonstrated with warm vapor and then cold thermal atoms \cite{hosseini_high_2011,cho_highly_2016}. Attractions of the GEM scheme, beyond high efficiency, have been the applications in information processing made possible by the bandwidth, frequency, and spatial manipulations allowed by the frequency gradient \cite{campbell_echo-based_2016}.

In this work, we show that the efficient operation of GEM relies significantly on the contribution of EIT. The mechanism that gives rise to this `EIT boost' can be understood by examining the absorption spectrum of the memory, shown in Fig.~\ref{fig:concept} a). The right hand blue peak is the 2-photon resonance, and produces coherent interaction of the light with the spinwave. The left hand orange peak is the resonant interaction, which causes scattering loss. To avoid this scattering, one can operate the memory at large detuning. Between these features there is a dip in absorption due to EIT resulting from destructive interference between the two interactions. Beyond the Raman peak, there is additional loss-causing absorption due to electromagnetically-induced absorption (EIA)\cite{lezama_electromagnetically_1999}.

GEM operates by changing the frequency of this 2-photon resonance along the length of the memory so that each frequency within the bandwidth of the signal pulse will be resonant somewhere in that length. These frequency components of light first travel through non-resonant atoms on their way to being absorbed, and then for a second time after being re-emitted. The sign of the frequency gradient determines whether the light travels through EIT or EIA, as shown in Fig.~\ref{fig:concept} c) and d). These conceptual diagrams illustrate how those frequency components interact with different parts of the absorption spectrum before arriving at their two-photon absorption peak.

The simplest way to operate GEM requires flipping the sign of the frequency gradient just once between storage and recall. The signal light continues through the memory, again travelling through a region of EIT or EIA, as in the recall sections of Fig.~\ref{fig:concept} c) and d). For one order of the applied gradients in c), negative then positive, the light travels through EIT twice, resulting in more efficient memory operation. For the order in d), positive then negative, the light travels through EIA twice, and the memory is less efficient. We call this dependence on the order in which the gradients are applied the \textit{gradient-order effect}.

%The interaction of light with the ensemble in GEM is illustrated in figure \ref{fig:concept} a). The blue absorption peak, at two-photon resonance in the atomic absorption spectrum, couples the light to the ensemble and stores it. An EIT window sits next to this in frequency space, where absorption of the light is reduced. When light travels through the atomic frequency gradient it can pass through this window before being absorbed in the memory, and there is reduced loss from incoherent absorption of the signal light. A larger, lossy absorption exists on the other side of the Raman absorption and causes increased loss to light passing through it. We equate this loss to the phenomenon of electromagnetically-induced absorption   The sign of the frequency gradient determines which of these regions the light passes through before being absorbed, and, because the gradient is typically flipped between storage and recall, as shown in figure \ref{fig:concept} b), light passes through the same side of the absorption peak while exiting the memory. The order in which gradients are applied determines whether light passes through EIT or EIA on its way into and out of the memory. We call the effect that this order has on efficiency the \textit{gradient-order} effect. We determine the size of this effect and its dependence on other parameters, test our predictions in experiments, and discuss its implications for applications of GEM.

The contribution of EIT to the memory efficiency is also dependent on the detuning. The EIT is reduced further off resonance and eventually disappears. The two-photon resonance then becomes a symmetric Lorentzian peak, removing the gradient-order effect. We use the two-photon absorption spectrum to determine how the effect depends on detuning, show there is an optimal detuning that depends on optical depth, and compare the theoretical form with simulations and experiments.

\section{\label{sec:level1}Theory}
The gradient-order effect occurs due to frequency-dependent interaction of light travelling through the memory. The size of  the EIT boost can therefore be extracted from the spectrum by separating it from the other important loss mechanisms. Those losses are the incomplete absorption of light by the memory, $L_{leakage}$, and resonant scatter of the stored light by the control field $L_{scatter}$. These do not directly depend on the detuning, so we quantify these at large detuning where there is no EIT boost. The losses correspond to the area and width, respectively, of the far-detuned Lorentzian two-photon absorption peak.  Once we have quantified the losses in the far detuned regime, we can make a comparison with the case where EIT reduces the loss, allowing us to quantify the significance of the EIT boost.

The absorption of the light, $\alpha(\omega)$, can be calculated from the imaginary part of the susceptibility  $Im[\chi(\omega)]$. For steady state populations, and strong control field, a signal field $\mathcal{E}(\omega)$ interacting with the 3-level $\Lambda$ system \cite{fleischhauer_electromagnetically_2005} will have absorption given by 
\begin{align}
\alpha(\omega)= \frac{d}{2}\left(\frac{8\delta^2\Gamma+2\gamma(|\Omega|^2+\gamma\Gamma)}{||\Omega|^2+(\Gamma+i2(\Delta_C+\delta))(\gamma+i2\delta)|^2}\right)\label{eq:susceptibility}
\end{align}

with optical depth $d$, signal detuning $\Delta_S = \omega-\omega_{13}$, control detuning $\Delta_C$, 2-photon detuning $\delta=\Delta_S-\Delta_C$, control field Rabi frequency $\Omega$, the 1-3 transition linewidth $\Gamma$, and dephasing rate $\gamma$ for the atomic coherence. For simplicity of the theoretical analysis we assume $\gamma=0$, although for numerical modelling $\gamma$ is set according to the experimental value. The total loss of the atomic coherence is generally small over the storage and recall of a single pulse. It is convenient for this analysis to work with the optical depth $d$ and normalise the ensemble length so that transmission $T(\omega) = \exp{(-\alpha(\omega))}$. At resonance and with no control field, $\alpha = d$ according to Eq.~\ref{eq:susceptibility}. At large detuning, the two-photon absorption becomes a Lorentzian, with width $\Gamma\Omega^2/\Delta^2$ and height $d$.

\begin{figure}
    \centering
    \includegraphics[width=\textwidth]{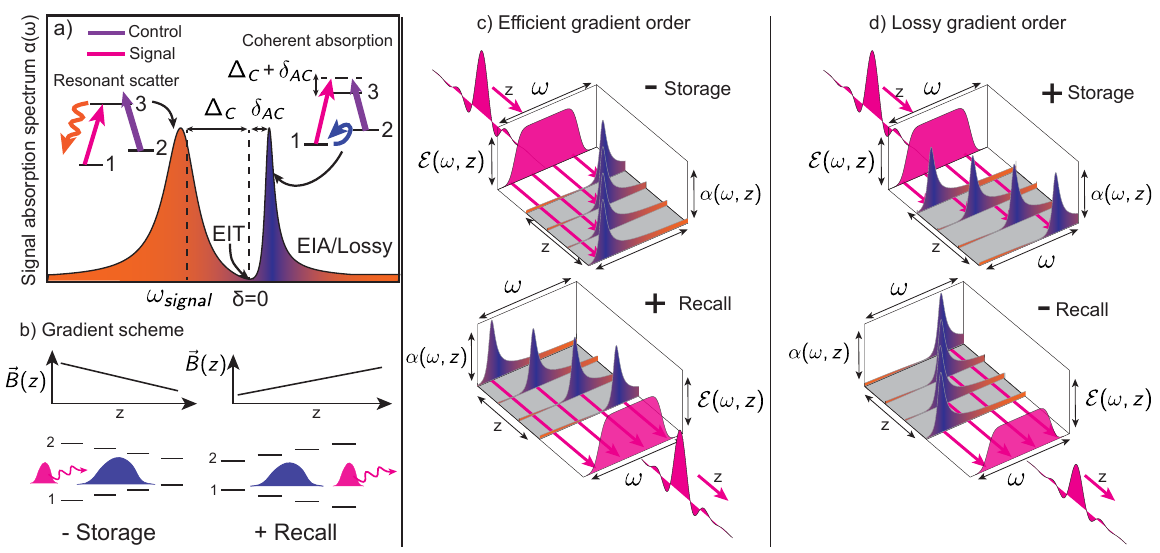}
    \caption{The frequency- and spatially-dependent interactions efficiency to depend on gradient order. a) The imaginary part of the susceptibility for the 3-level scheme, shown with a very small detuning to exaggerate the EIT and EIA. b) A magnetic field gradient causes the 2-photon detuning and so the coherent absorption frequency to change with position. c) A pulse of light (magenta), plotted in z (not to scale),  enters the memory (grey). The various frequency components, plotted in $\omega$, travel through the memory until they are coherently absorbed, passing through EIT on their way. The gradient is flipped to recall the light, which again travels through EIT while exiting the memory. d) Reversing the gradient order makes the memory less efficient, as the light travels through EIA before being stored, and again after being retrieved. The gradient is continuous and linear, not piecewise, but only slices of the spectrum are shown to make the absorption magnitude more obvious.}
    \label{fig:concept}
\end{figure}

An absorption spectrum illustrative of GEM is plotted in Fig.~\ref{fig:concept} a),
representing the equilibrium atom-light interaction for light at each particular signal detuning. The interaction of light with the memory is not quite at equilibrium due to the short pulse times, but the spectrum still roughly corresponds to different features of the atom-light interaction in the memory. The left peak, coloured orange, is resonant scatter and loss of the signal light, which is avoided by using a sufficient control detuning $\Delta_C$. The narrower peak, coloured blue, is coherent absorption into the spinwave.

EIT and EIA can be identified with a third feature - an asymmetry in the coherent absorption line due to the finite detuning. This is an antisymmetric dispersion line that has the same magnitude as the tail of the resonant absorption, since on the EIT side it cancels exactly to give $\alpha(\delta=0)=0$, or complete transparency. The absorption spectrum that light interacts with while being stored and recalled from the memory, and its dependence on the gradient, is illustrated in Fig.~\ref{fig:concept} c) and d). The paths the frequency components take through lossy or transparent regions while travelling to the absorption peak, and then through to the end of the memory upon recall are shown by magenta arrows. In order to make the EIA and EIT visible in these plots, the detuning is set too low for efficient GEM, but the spectra are otherwise accurate for GEM operation.

 Now, we turn to the optimal operation of the memory to allow us to quantify the gradient-order effect. At far detuning ($\Delta_C \gg \Gamma$) only the leakage of light through the memory during storage, $L_{leakage}$, and the control field resonant scatter, $L_{scatter}$, contribute to memory inefficiency.  To maximise the memory storage and recall, which are symmetric processes, we must therefore maximise $(1-L_{leakage})(1-L_{scatter})$.
 
 In the far detuned regime, Eq.~\ref{eq:susceptibility} gives a Lorentzian Raman absorption peak. When a gradient of bandwidth $BW$ is applied, the Raman absorption is spread across the bandwidth, which we assume is significantly larger than the unbroadened Raman line. To find the absorption of the signal by this broadened ensemble we divide the area under the Lorentzian by the bandwidth, giving the amount of leakage as
\begin{align}
L_{leakage}& = \exp{\left(\frac{- \pi   \Gamma \Omega^2d}{BW\Delta_C^2} \right)}. \label{eq:leak}
\end{align}

 To minimise the scattering due to the control field, we apply it only for the duration required, which for a Fourier limited signal pulse will be $2\pi/BW$, where we have also assumed that the bandwidth of the memory ($BW$) has been matched to the bandwidth of the signal. Light at the leading edge of the signal is subject to the whole period of the control pulse, while light stored last will have negligible exposure to the control field, so the scattering, averaged across the pulse, drives the signal field for half the control field duration. Recalling that that it is the width of the Raman absorption that is used to find the scattering loss we obtain 
\begin{align}
L_{scatter} & = 1 - \exp{\left(\frac{-\pi\Gamma\Omega^2}{BW\Delta_C^2} \right) \label{eq:controlscatter}}
\end{align}

Maximizing $(1-L_{leakage})(1-L_{scatter})$ gives
\begin{align}
\frac{\pi\Gamma\Omega^2}{BW\Delta_C^2}&=\frac{\log 2}{d}.\label{eq:optimalparams}
\end{align}

In this regime, the total loss only depends on $d$, since any of the other parameters can be adjusted with respect to each other while still achieving the same total efficiency. For example, increasing the bandwidth ($BW$) or detuning ($\Delta_C$) would require a larger control field Rabi frequency ($\Omega$) to maintain the efficiency, but the same maximum efficiency could be reached.

The far-detuned regime will be used as a benchmark for comparison to the regime where EIT and EIA play a role, so that we can assess the impact they have on memory performance. We therefore define $\alpha'(\omega)$ to be the difference between the far detuned Lorentzian Raman absorption spectrum and the asymmetric near-detuned Raman spectrum. The degree to which EIT and EIA impact the memory performance will be captured by integrating $\alpha'(\omega)$ over the memory bandwidth.  So that we correctly account for the gradient, which is flipped for storage and recall, we in fact integrate $2\alpha'(\omega)$ over half the memory bandwidth, where the choice of which half determines whether the signal experiences EIT or EIA. $\alpha'(\omega)$ is antisymmetric around the center of the memory, except for an additional constant term at small detuning, so the EIT and EIA produce equal and opposite efficiency gain or loss. 

Leaving $d$ and $\Delta_C$ as free parameters and constraining the other parameters according to equation \ref{eq:optimalparams}, the gradient-order effect $G$ is, approximately

\begin{align}
    G&=\pm \int_{BW/2}-2\alpha'(\omega)d\omega\\&\approx\frac{d}{2\pi \Delta_\Gamma^2}+\log(4)\left(1+\frac{2}{\pi\Delta_\Gamma}\left(\log{\left(\frac{\log{(4)}}{ d}\right)}-\Delta_\Gamma\arctan{(2\Delta_\Gamma)}\right)\right).\label{eq:effectmagnitude}
\end{align}
for $d,\Delta_\Gamma=\Delta_C/\Gamma \gg 1$. This is an integrated absorption, so we expect the efficiency of the near-detuned memory, $\eta_N$, compared to that of the far-detuned memory, $\eta_F$, to be approximately $\eta_N=\exp(G)\eta_F$, where a positive value of $G$ in the exponent corresponds an increased efficiency due to EIT, and a negative value corresponds a reduced efficiency due to EIA.  

\begin{figure}
    \centering
    \includegraphics[width=0.9\textwidth]{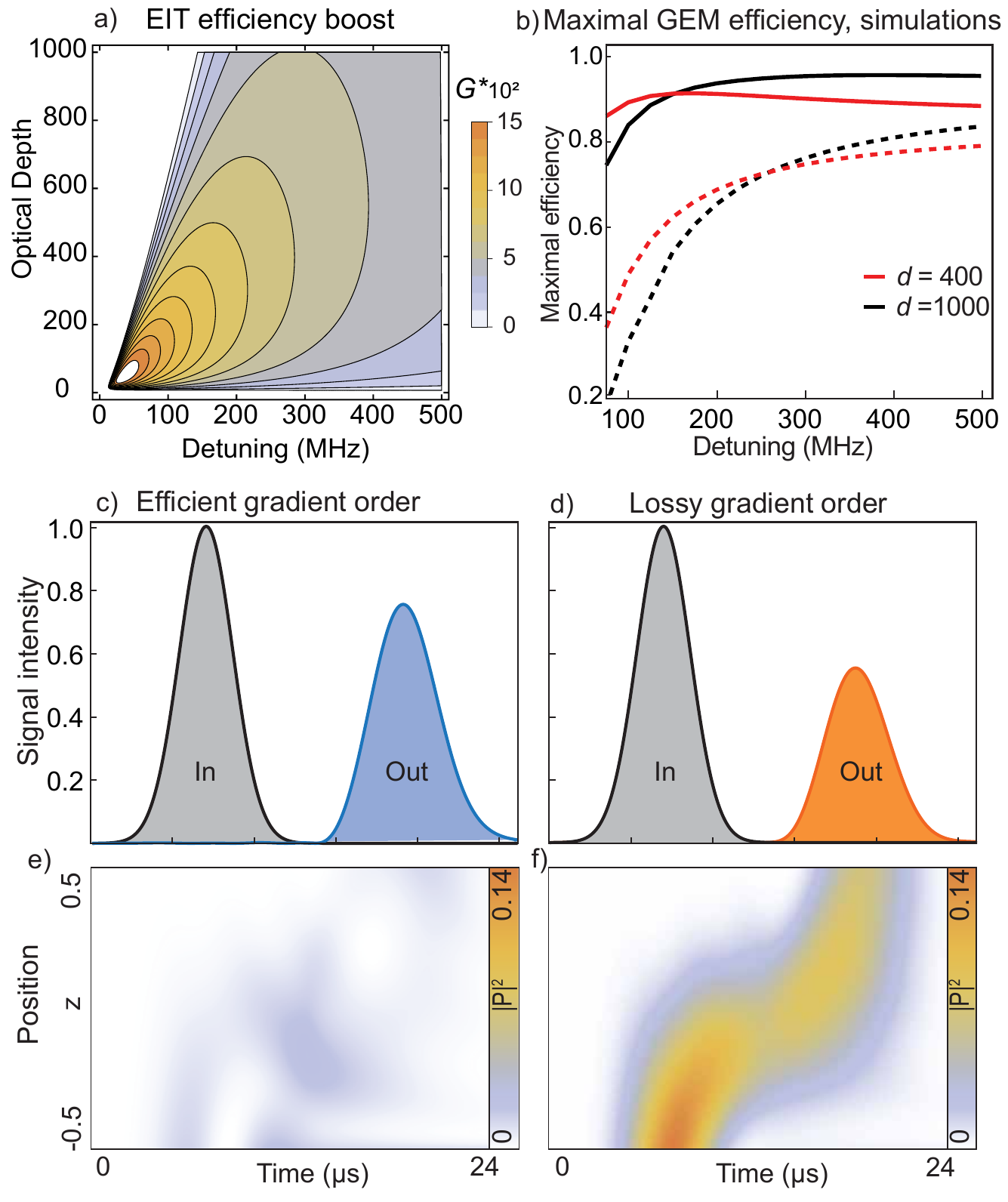}
    \caption{a) Amount of extra memory efficiency due to the EIT boost, approximated by integrating the EIT window within the memory bandwidth. b) Maximal simulated efficiency of pulse storage and recall at two different optical depths and detunings. Optical depths of 400 (red lines) and 1000 (black lines), storing with a decreasing (solid) or increasing (dashed) frequency gradient and then retrieving with the opposite gradient. c-f) Sample XMDS simulation results for optical depth of 400 and detuning of 180 MHz. The population of the excited state for the efficient order, e) is greatly reduced compared to the lossy order in f).}
    \label{fig:theory}
\end{figure}

Setting $\Gamma=2\pi*5.75$ MHz for the memory transition used in simulation and experiment, $G$ is plotted in Fig.~\ref{fig:theory} a) for relevant detunings and optical depths.  This plot shows that in the limit of large detuning, there is no EIT boost.  As the detuning is reduced, however, the analytic model shows that there is a reduction in the amount of loss experienced by light and that for a given optical depth, there is an optimal detuning that is roughly proportional to the optical depth.  

The analytic model provides some insight into the dependence of the gradient order effect on the memory parameters and an estimate of how strong an effect it could be. To assess more precisely what happens in the GEM protocol, we turned to numerical simulations. We used XMDS \cite{dennis_xmds2:_2013} and the following equations to test the relationship between optical depth, detuning, and efficiency. 
\begin{align}
\partial_tS& =i(BW z-\gamma)S +  \sfrac{i\Omega}{2} P\\
\partial_tP& =i(\Delta_C-\sfrac{\Gamma}{2})P +  \sfrac{i\Omega}{2} S + \sfrac{ i\Gamma}{2} \sqrt{\sfrac{d}{2}} \mathcal{E}\\
\partial_z\mathcal{E}& =i\sqrt{\sfrac{d}{2}} P \label{eq:polarization}
\end{align}

A derivation of these equations can be found in Appendix A of \cite{gorshkov_photon_2007}. That work emphasizes the importance of optical depth as the key resource for atom-optic memory efficiency, which also informs our analysis.

The incoming signal light is applied as a LHS boundary condition of  $\mathcal{E}(t)$, and the outgoing light is measured at the RHS. $S$ and $P$ represent coherent excitations of the atoms, scaled so that optical depth $d$ can be used to represent the interaction strength of the signal light. The spinwave $S$ is a coherence between levels 1 and 2 in Fig.~\ref{fig:concept} (a), and $P$ is a coherence between levels 1 and 3. $P$ is proportional to the polarization of the atoms, hence the differential equation \ref{eq:polarization}. We also use these simulations to analyse the experimental results by setting a nonzero loss, $\gamma$, consistent with losses in the experimental system due to atomic motion.

For two different optical depths and a constant pulse bandwidth, we set the memory bandwidth and control field Rabi frequency to maximize the efficiency to store and recall a single Gaussian pulse. We varied the detuning in 25 MHz increments to show that the efficiencies for opposite gradient orders converge slowly as the detuning increases, with results provided in Fig.~\ref{fig:theory} b). There is a maximum efficiency at a detuning roughly proportional to the optical depth. For $d={400, 1000}$, the maximum simulated efficiencies are $\eta={0.914, 0.945}$ at $\Delta_C= 175,375$ MHz, as predicted by the parametric analysis. The drop in efficiency beyond this detuning is most significant at lower optical depths, and becomes smaller for larger optical depths. The efficiency of the lossy gradient order monotonically increases over the range we simulated, and should converge with the efficient order at very large detunings. 

Equation \ref{eq:effectmagnitude} closely predicts the optimal detuning and the ratio between peak and far-detuned efficiency, which should be equal to the exponential of the integral. For example, the highest simulated efficiency for $d=400$, at $\Delta_C=175$ MHz achieved $\eta_N=0.914$. We approximate $\eta_F \approx 0.83$ by averaging the far-detuned simulated efficiencies. Using equation \ref{eq:effectmagnitude} to obtain $G=0.08$ gives $\exp(G)\eta_F\approx0.9$.

\section{Experiment}
\begin{figure}
    \centering
    \includegraphics[width=\textwidth]{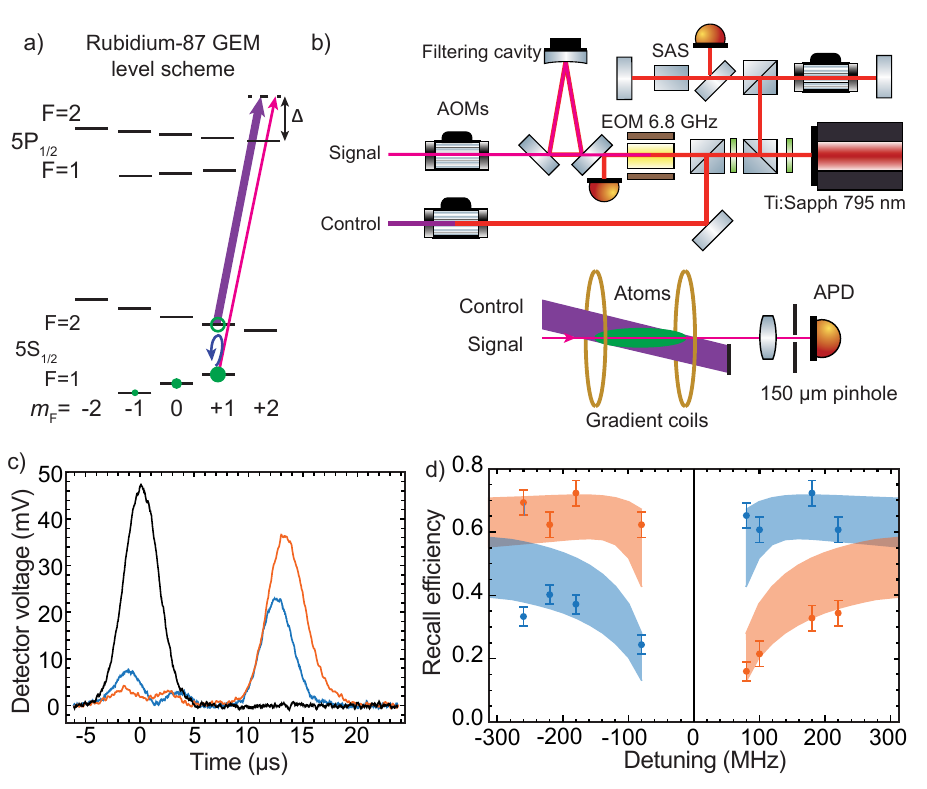}
    \caption{a) The level scheme used for this GEM experiment. A 2-photon interaction between the signal (magenta) and the control (purple) maps the signal into a spinwave between the first state and the $\mathrm{F}=2, m_\mathrm{F}=+1$ level. b) The control and signal are derived from a single Ti:Sapph laser, locked using  Zeeman-modulated saturated absorption spectroscopy (SAS). AOMs and a fiber EOM are used to reach the detunings selected.  c) Example photo-detector trace comparing storage plus recall between efficient and lossy gradient orders, taken at -180 MHz detuning and averaged over 50 repetitions of the experiment. d) Recall efficiencies after 13 $\upmu$s storage, plotted against simulated efficiencies (shaded areas) for an optical depth of 400. Negative gradient followed by positive (blue), and positive followed by negative (orange). The swapping of the colours between negative and positive detunings means that the efficient gradient order was reversed.}
    \label{fig:experiment}
\end{figure}

We used an ensemble of cold rubidium-87 atoms generated in the setup described in Cho et al. \cite{cho_highly_2016}. The atoms were magneto-optically trapped, compressed, cooled and optically pumped to $\mathrm{F}=1$, $m_\mathrm{F}=+1$ to produce an ensemble with optical depth 450 $\pm$50 on the signal transition of Fig.~\ref{fig:experiment} a). Lasers to prepare the atomic ensemble were derived from external-cavity diode lasers locked to saturated absorption spectra (SAS), amplified in tapered amplifiers and frequency-amplitude-controlled with acousto-optic modulators (AOMs). Control and signal were derived from a Ti:sapph laser locked to a SAS, with frequency and amplitude controlled via a fiber electro-optic modulator (EOM) and AOMs, see Fig.~\ref{fig:experiment} b). Coils in anti-Helmholtz configuration produced the magnetic field gradient along the atomic ensemble. The signal was sent through the long axis of the ensemble, with the control overlapping at a small angle throughout. The signal was spatially filtered with a 100 $\upmu$m pinhole to remove the control laser before the detector. The experiment was run over a range of control field detunings $\Delta_C$ of $\pm$260 MHz. A spatially homogeneous bias magnetic field separates the 2-photon transitions from adjacent $m_\mathrm{F}$ levels by about 1 MHz, and a frequency gradient of 300 $\pm$50 kHz was applied using the gradient coils. Uncertainties in optical depth and gradient are due to the spatially inhomogeneous atom ensemble.

A 5 $\upmu$s FWHM Gaussian pulse was stored and recalled, and the control field intensity, memory bandwidth, and signal carrier frequency were adjusted to maximize the recall efficiency. A range of control field detunings were used to allow comparisons with our modelling.

Efficiency is the integral of output pulse energy divided by input pulse energy, as measured at an avalanche photo-detector after the atoms. The atoms were removed to measure the input pulse without absorption by the atoms. The results are plotted in Fig.~\ref{fig:experiment} d). When data was taken for a series of detunings, the control field intensity was adjusted for each detuning, and then the carrier frequency of the pulse was changed to account for the AC-Stark shift of the memory's central frequency. The low SNR for the inefficient gradient in particular makes it difficult to set the control field intensity accurately.

We tested the experimental results against simulations run with the same optical depth, bandwidth, and with optimal control Rabi intensities. It was not possible the experimental results directly to equation \ref{eq:effectmagnitude} due to the a dephasing $\gamma\neq0$. Spatial variation of the control field intensity and atoms in other $m_\mathrm{F}$ levels added some additional losses that could not be accurately modelled. Instead, to approximately account for these losses, we ran simulations at a sub-optimal setting of 20\% higher $\Omega$.  We plot coloured areas between these bounds to give a comparison to the experimental data.

\section{Discussion}
The experimental results agree with the theory, and the different efficiencies of the two gradient orders converge as the detuning is increased. The signs of the gradients that give the lossy and efficient gradient orders are swapped when the sign of the detuning is switched. Our model also predicts a peak in the efficiency at moderate detuning. Our experiment was not able to definitively show this behaviour owing to difficulties in running the experiment at large detuning.

Obtaining high efficiency at large detunings is experimentally challenging, as a spatially inhomogeneous control field causes inhomogeneous phase shifting of the spinwave via AC-Stark shift. Variation in the spinwave phase transverse to the wavefront of the light causes steering and defocussing of the recalled light, reducing the efficiency of the spatially filtered echo. The AC-Stark shift scales with control field intensity divided by detuning ($\propto|\Omega|^2/\Delta_C$), and considering the control field intensity for a given optical depth, detuning, and bandwidth, we obtain a scaling $\delta_{AC} \propto d/\Delta_C$. Since the gradient-order effect gives the best efficiency at $d \propto \Delta_C$, we expect the reduced efficiency observed at higher detuning is due to a combination of the phase shift and the gradient-order effect. The supplementary material includes simulation data quantifying this effect, showing that its impact on the efficiency at larger detuning is indeed similar to the predicted gradient-order effect. This makes it difficult to separate the two effects and conclusively establish a drop in efficiency at larger detunings due to the gradient-order effect.

Figure \ref{fig:experiment}. c) shows averaged traces comparing the detected output pulses for the opposite gradient orders. A remarkable feature is the earlier recall using the lossy order. This is due to a slow light effect caused by the transparency window in the efficient order, and a fast light effect caused by the extra loss in the lossy order. Recall timings can easily be adjusted in GEM, but this was an unexpected demonstration of the dispersive part of the gradient-order effect.

The lossy side of the effect is due to EIT's counterpart, EIA. EIT exists due to a destructive interference between multiple excitation pathways of the atoms, and schemes to produce EIA generally introduce additional excitation pathways to disrupt or modify this interference \cite{das_interplay_2019}. In the case of GEM, no additional pathways are introduced to obtain EIA, and the interference becomes destructive or constructive due to the spatial gradient in the atomic frequency.

High efficiency GEM was demonstrated over a decade ago in warm vapor \cite{hosseini_high_2011} and more recently in cold thermal atoms \cite{cho_highly_2016}. The gradient-order effect was previously observed when setting up experiments, but without a theoretical explanation it was believed to be caused by experimental imperfections. The previous high efficiency results would have been taken with the efficient gradient order, as the lossy order has a significantly lower theoretical efficiency than the experimentally measured efficiency.

An interesting prospect to consider is that, outside of GEM, memory schemes based on EIT and other interactions are also subject to losses intrinsic to the absorption spectrum of the three-level interaction. We are not aware of a way to take advantage of an asymmetric spectrum without a spatial gradient in frequency, but perhaps this detailed explanation of its impact on GEM will be useful in future discoveries. 

\section{Acknowledgements}
We thank Geoff T. Campbell for his contributions to earlier work on this project.

\bibliography{main}
\end{document}